\documentclass[twocolumn]{bmcart}
\usepackage{graphicx}
\usepackage{ragged2e}
\usepackage{algorithm}  
\usepackage{algorithmic}

\usepackage{amsthm,amsmath}
\usepackage[utf8]{inputenc}

\startlocaldefs
\endlocaldefs

\begin{document}

\begin{frontmatter}

\begin{fmbox}
\dochead{Research}
\author[
  addressref={aff1},
  corref={aff1},
  email={lxinrui10@gmail.com}
]{\inits{X.R.L.}\fnm{Xinrui} \snm{Liu}}
\author[
  addressref={aff1},
  email={wangyajie0312@foxmail.com}
]{\inits{Y.J.W.}\fnm{Yajie} \snm{Wang}}
\author[
  addressref={aff1},
  email={tan2008@bit.edu.cn}
]{\inits{Y.A.T.}\fnm{Yu-an} \snm{Tan}}
\author[
  addressref={aff1},
  email={kfqiu@bit.edu.cn}
]{\inits{K.F.Q.}\fnm{Kefan} \snm{Qiu}}
\author[
  addressref={aff1},
  email={popular@bit.edu.cn}
]{\inits{Y.Z.L.}\fnm{Yuanzhang} \snm{Li}}

\address[id=aff1]{%
  \orgdiv{School of Cyberspace Science and Technology},
  \orgname{Beijing Institute of Technology},          
  \city{Beijing},                              
  \cny{China} 
}


\title{Stealthy Low-frequency Backdoor Attack against Deep Neural Networks}

\begin{abstractbox}

\begin{abstract} 
\justifying
Deep neural networks (DNNs) have gain its popularity in various scenarios in recent years. However, its excellent ability of fitting complex functions also makes it vulnerable to backdoor attacks. Specifically, a backdoor can remain hidden indefinitely until activated by a sample with a specific trigger, which is hugely concealed. Nevertheless, existing backdoor attacks operate backdoors in spatial domain, i.e., the poisoned images are generated by adding additional perturbations to the original images, which are easy to detect. To bring the potential of backdoor attacks into full play, we propose low-pass attack, a novel attack scheme that utilizes low-pass filter to inject backdoor in frequency domain. Unlike traditional poisoned image generation methods, our approach reduces high-frequency components and preserve original images' semantic information instead of adding additional perturbations, improving the capability of evading current defenses. Besides, we introduce "precision mode" to make our backdoor triggered at a specified level of filtering, which further improves stealthiness. We evaluate our low-pass attack on four datasets and demonstrate that even under pollution rate of 0.01, we can perform stealthy attack without trading off attack performance. Besides, our backdoor attack can successfully bypass state-of-the-art defending mechanisms. We also compare our attack with existing backdoor attacks and show that our poisoned images are nearly invisible and retain higher image quality.
\end{abstract}


\end{abstractbox}

\end{fmbox}

\end{frontmatter}

\section*{Introduction}
\label{sec:intro}
\begin{figure}
  \centering
  \includegraphics[width=0.9\linewidth]{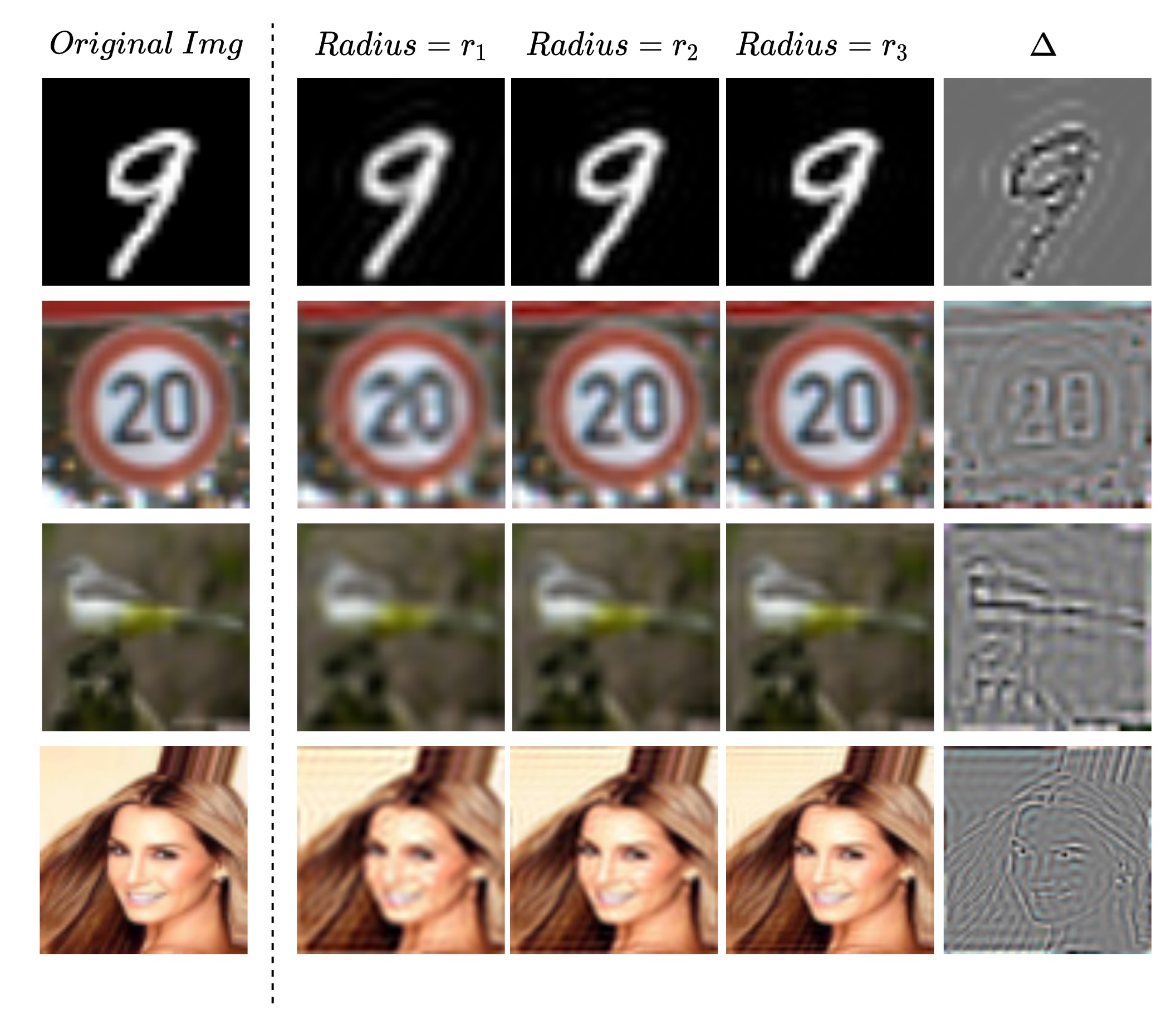}
  \caption{The figure shows original images, poisoned images with three different low-pass radiuses ($r_1<r_2<r_3$) and corresponding residual maps. }%
  \label{fig:dataset_radius}
\end{figure}
\begin{figure*}[tbp]
    \centering
    \includegraphics[width=2\linewidth]{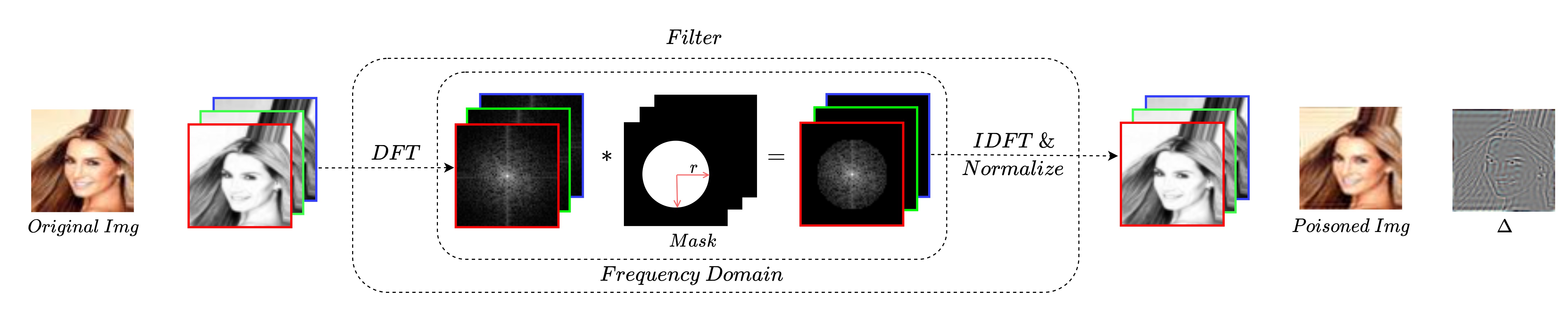}
    \caption{Pipeline of generating low-pass based poisoned images. $r$ denotes the low-pass radius. Note that low-frequency components are clustered in the center of the frequency-domain image.}%
    \label{fig:low_pass_pipeline}
\end{figure*}
Serving as the backbone, neural networks are increasingly playing  important roles in promoting advanced artificial intelligence applications. Currently, neural networks have been adopted in a variety of applications, including face recognition~\cite{russakovsky2015imagenet}, voice recognition~\cite{graves2013speech}, games~\cite{hermann2013multilingual}, and autonomous driving~\cite{bojarski2016end}. However, despite great advantages, recent studies have shown that deep learning models are vulnerable to backdoor attacks~\cite{gu2017badnets}.\par

Backdoor attack enables an adversary to implant a backdoor into the original model and initiate attack via a specific backdoor trigger. The backdoored deep neural network can correctly classify benign samples but misclassify any input with a specific backdoor trigger. The backdoor usually maintain hidden until activated by a specific backdoor trigger, which is hugely concealed. Therefore, it cause serious security risks in many applications.\par

However, current backdoor attacks all assume that the poisoned image is generated by adding additional perturbations to the original one in spatial domain.

Unfortunately, adding patch~\cite{gu2017badnets}, blending~\cite{chen2017targeted} and reflection~\cite{liu2020reflection} to the image can be easily detected by existing defense methods~\cite{wang2019neural}~\cite{gao2019strip}~\cite{liu2018fine}. 
To solve this problem, we apply filter to generate poisoned images, which can filter out high-frequency noise and preserve original images' semantic information instead of adding extra perturbations, thus passing current defenses easily. In multiple filter algorithms, low-pass filter~\cite{motwani2004survey} allows a specific range of low-frequency information to be preserved. As explained in ~\cite{wang2020high}, low-frequency components have major semantic information and almost look identical to the original image to human.

In this paper, we propose a novel and effective backdoor attack through frequency domain, termed low-pass attack. Specifically, figure~\ref{fig:low_pass_pipeline} demonstrates the pipeline of generating a poisoned image. First, we convert the original image from spatial domain to frequency domain. Low-frequency components are clustered in the center of frequency-domain image. Given the attacker chosen radius: $r$, we design a "Mask" to preserve specified range of information. The larger the low-pass radius: $r$, the more original low-frequency components are retained in the poisoned image. Then we can transform the filtered frequency-domain image back to spatial domain to generate a poisoned image. As shown in figure~\ref{fig:dataset_radius}, the residual map properly reflects the contour information of the original image, making Our poisoned images natural and indistinguishable from the original inputs. \par

To obtain a backdoored model, we first use radius: $r_{t}$ to filter a part of training images and modify corresponding labels for training. Though our trained network provide high attack performance, our evaluation shows that images with low-pass radiuses: $r^{'}$($r^{'}\neq r_{t}$ but close to $r_t$) will also be misclassified as the target with a high success rate. This makes our attack lose some stealthiness since the backdoor can be triggered as long as the low-pass radius is within a certain range. Hence, to improve the stealthiness of our attack, we propose "precision mode", which makes our backdoor only be precisely triggered by a specified low-pass trigger radius: $r_{t}$.\par

We evaluate our low-pass backdoor attack on four datasets, including MNIST, GTSRB, CIFAR10 and CelebA. The experimental results demonstrate that even under pollution rate of 0.01, we can perform stealthy backdoor attack without compromising attack success rate and clean sample accuracy. Besides, our attack is undetectable by three popular defense mechanisms, including STRIP, Fine-pruning and Neural cleanse. We also compare our attack with several existing backdoor attacks and show that our poisoned images are nearly invisible and retain higher image quality. \par
The contributions of this paper are as follows:\par

\begin{itemize}
\item We propose a novel backdoor attack through frequency domain, termed low-pass attack.
\item We introduce "precision mode" to train our model, allowing the backdoor to be triggered precisely and stealthily.
\item Empirical results show that our attack is effective, stealthy and robust against existing defenses.
\end{itemize}

\section*{Related work}
\label{sec:related_work}

\subsection*{Backdoor attack}
Backdoor attacks first took effect in neural networks in 2017. Gu et al.~\cite{gu2017badnets} proposed BadNets. In this work, the attacker can attach a specific trigger to the stop sign image and mark it as the speed limit sign to generate a backdoor in the road sign recognition model. Although the model can correctly classify clean samples, it will misclassify the stop sign image with the trigger as the speed limit.\par
In 2018, Liu et al.~\cite{liu2017trojaning} proposed a more advanced backdoor attack technique called Trojan attack. In the study of the Trojan attack, it was found that the backdoor attack method in the neural network was effective because the backdoor trigger would activate specific neurons in the network. Therefore, the Trojan attack generates a trigger in a way that maximizes the activations of specified neurons. In 2019, Yao et al. ~\cite{yao2019latent} introduced backdoor attack into transfer learning, the backdoor was stubborn and could remain active after transferring from teacher model to student model. Besides, many recent studies focus on clean-label backdoor attack which can exhibit similar unexpected behavior on specific inputs, but the adversary can inject backdoor on poisoned samples with unchanged labels to evade human inspections. \par
Currently, most of the backdoor attacks depend on fixed patch based trigger. Salem et al.~\cite{salem2020dynamic} systematically studied the praticality of dynamic trigger, which can hijack the same label with different trigger patterns and different trigger locations. Chen et al. ~\cite{chen2017targeted} generated poisoned images by blending the backdoor trigger with benign inputs rather than stamping the trigger. Liu et al.~\cite{liu2020reflection} propose to implant backdoor with a delicate crafted natural trigger, namely refection, which is a common natural phenomenon.\par
\subsection*{Backdoor defense}
Research on backdoor attacks on neural networks has also prompted corresponding defense research and has become a new research field. Multiple methods have been proposed to defend against backdoor attacks.\par
As an earlier defense, Liu et al.~\cite{liu2018fine} propose to remove potential backdoor by firstly pruning carefully chosen neurons of the DNN model that contribute least to the classification task. After pruning, fine-tuning is used to restore model performance. In 2019, Neural Cleanse~\cite{wang2019neural} was proposed. It optimized a patch-based trigger candidate for each label and then detected if any candidate was abnormally smaller than the others. Unlike model inspection mentioned above, STRIP~\cite{gao2019strip} is a test time defense. It takes advantage of the feature that any input containing a backdoor will be classified as the target. The superimposed poisoned data will still be misclassified, but the output of superimposed clean data is relatively random. The method distinguishes clean inputs from poisoned inputs by determining the information entropy. In contrast, NEO~\cite{udeshi2019model} searched for the candidate trigger patches where region blocking changed the predicted outputs.

\section*{Methodology}

\subsection*{Threat model}
We assume a user who wants to use a training dataset $D_{train}$ to train the parameters of a DNN. The user sends the internal structure of the DNN $M$ to the trainer. Finally, the trainer will return to the user the trained model parameters $\Theta^{'}$. However, the user cannot fully trust the trainer. The user needs to check the accuracy of the trained model on the validation dataset $D_{valid}$. Only when the model's accuracy meets an expected accuracy rate $a^*$ will the user accept the model.
\subsection*{Attacker's Goals}
The attacker expects to return to the user a maliciously trained backdoor model parameters $\Theta^{'}:=\Theta^{adv}$. The parameters of this model are different from those of the honestly trained model. A backdoored model needs to meet two goals:\par
Firstly, the classification accuracy of the backdoored model $M_{{\Theta^{adv}}}$ cannot be reduced on the validation set $D_{valid}$, in other words, that is, $C(M_{\Theta^{adv}},D_{valid})\ge a^*$. Note that the attacker cannot directly access the user's validation dataset.\par
Secondly, for the input containing the backdoor trigger specified by the attacker, $M_{\Theta^{adv}}$ outputs' predictions are different from the outputs of the honestly trained model.\par

\subsection*{Generate low-pass based poisoned images}

In canonical backdoor trigger design approaches, the backdoor triggers are all based on noise perturbations, making poisoned data easily detected by current backdoor defenses. In this paper, we apply low-pass filter in image denoising field to filter out certain high-frequency features to generate poisoned images. As shown in figure~\ref{fig:low_pass_pipeline}, to control the denoising degree of poisoned images, we define a low-pass radius: $r$ to control the amount of the passed low-frequency features. The larger the low-pass radius, the more low-frequency features remain, and vice versa.

\begin{align}
    \begin{split}
    F(u,v)&=DFT(f(p,q))\\&=\sum_{p=0}^{H-1}\sum_{q=0}^{W-1}f(p,q)e^{-i2\pi(\frac{up}{H}+\frac{vq}{W})}
    \end{split}
    \label{Equation:DFT}
\end{align}
\begin{align}
    \begin{split}
    f(p,q)&=IDFT(F(u,v))\\&=\frac{1}{HW}\sum_{u=0}^{H-1}\sum_{v=0}^{W-1}F(u,v)e^{i2\pi(\frac{up}{H}+\frac{vq}{W})}
    \end{split}
    \label{Equation:IDFT}
\end{align}\par

\begin{algorithm}[tb]
\caption{Low-pass Attack}
\label{alg:algorithm}
\raggedright
\textbf{Input}: Original model's internal structure: $M$; Original training images: $X$ and its corresponding label set: $Y$; Original training set: $D_{train}=(X,Y)$; Attack target: $t$ \\
\textbf{Hyper-parameters}: Pollution rate: $\beta$, Attacker chosen trigger radius: $r_{t}$, Trigger radius neighborhood: $\delta$ \\
\textbf{Output}: Retrained model's parameter: $\Theta^{adv}$\\

\begin{algorithmic}[1] 
\STATE Select $\beta*D_{train}$ as poisoned dataset $D_{poisoned}$ and precision dataset $D_{precision}$,  $(1-\beta)*D_{train}$ as clean dataset $D_{clean}$.
\STATE Normalize $\forall (x_{i}, y_{i}) \in D_{clean}$ to $[0,1.0]$.
\FOR{$(x_{i},y_{i})$ in $D_{poisoned}$}
    \STATE $x_{i}:=\mathcal{A}(x_{i}, r_{t})$\\
    \STATE $y_{i}:=t$
\ENDFOR

\FOR{$(x_{i},y_{i})$ in $D_{precision}$}
    \STATE $r_{p}=RandInt([r_{t}-\delta, r_{t})\ \& \ (r_{t}, r_{t}+\delta])$
    \STATE $x_{i}:=\mathcal{A}(x_{i}, r_{p})$\\
    \STATE $y_{i}:=y_{i}$
\ENDFOR

\STATE Retrain target classifier parameter.\\
    $\Theta^{adv}\leftarrow D_{clean}+D_{poisoned}+D_{precision}$\\
\STATE \textbf{return} $\Theta^{adv}$
\label{algorithm:Low-pass_attack_precision}
\end{algorithmic}
\label{algorithm:Low-pass_attack_precision}
\end{algorithm}

We assume that we have a grayscale image that can be viewed as an $H*W$ matrix ($H$, $W$ denote the height and width of the image, respectively). We can regard this image as a signal $f(p,q)$ ($f(p,q)$ denotes the pixel value of the spatial domain image at the coordinate point $(p,q)$). We utilize Discrete Fourier Transform (DFT) to convert an image from spatial domain to frequency domain. Here, we apply $F(u,v)$ to denote the pixel value of an image in frequency domain at the coordinate point $(u,v)$. Equation~\ref{Equation:DFT} represents Discrete Fourier Transform, and Equation~\ref{Equation:IDFT} represents the Inverse Discrete Fourier Transform (IDFT), which transforms an image from frequency domain to spatial domain. Note that $i$ denotes a unit of the complex number. 

In our generation procedure, we first convert the original image $f_{original}$ to frequency domain using DFT (Equation~\ref{Equation:DFT}), the output is represented as $F_{original}$. Then, we generate a low-pass mask $F_{mask}$ in frequency domain according to the attacker chosen low-pass radius $r$. The larger the low-pass radius, the more low-frequency features remain, and vice versa. The mask is essentially a matrix containing only $0$ and $1$. As shown in figure~\ref{fig:low_pass_pipeline}, the pixel value is $1$ within the radius region and $0$ outside the radius region. After that, we multiply the corresponding pixel values in $F_{original}$ and $F_{mask}$ to generate $F_{poisoned}$.
Finally, we convert the poisoned image in frequency domain back to spatial domain by performing IDFT (Equation~\ref{Equation:IDFT}). $f_{poisoned}$ is our generated space-domain poisoned image. In order to simplify the procedure above, we define a function $\mathcal{A}$ to represent the process of converting $f_{original}$ to $f_{poisoned}$. Given an image: $x$ and a chosen radius: $r$, we can obtain a low-pass filtered image $x^{\prime}$:
\begin{equation}
\begin{aligned}[b]
    x^{\prime} = \mathcal{A}(x, r)
    \label{Equation:convert}
\end{aligned}
\end{equation}

For RGB image, we split the image into three channels. For each channel, we generate a low-pass poison sample. The shape of the low-pass mask is $H*W*3$. The magnified residual map is shown in the fifth column in figure~\ref{fig:dataset_radius}. Our generated poisoned images are identical to the original images and the residual map is undetectable.

\subsection*{Boost attack with precision mode}
After applying an attacker chosen low-pass radius: $r_{t}$ to train a backdoored model, we found images with low-pass radius: $r^{\prime}$($r^{\prime}\neq r_{t}$ but close to $r_t$) would also be misclassified as the target with a high attack success rate in the inference phase. This phenomenon makes our attack lose some stealthiness as the backdoor can be triggered as long as the low-pass radius:$r^{\prime}$ is within a certain range. To improve the stealthiness of our attack, we propose "precision mode" so that our backdoor can only be precisely triggered by an attacker chosen low-pass radius:$r_{t}$. \par

We divide our low-pass attack into three different modes. Equation~\ref{Equation:precision_loss} illustrates ways to convert the same data pair(image, label) under three different modes.
Clean mode is used to maintain the functionality of original model. Attack mode enables the backdoor to be triggered by specified trigger radius:$r_t$. As for precision mode, we assume an adversary wants to trigger a low-pass radius: $r_{t}$. $r_{max}$ denotes the max radius of original image, that is, $\mathcal{A}(x, r_{max})$ and $x$ are equivalent. We define $\delta$ as the neighborhood of the trigger radius $r_{t}$. For each image $x_{i}$ using precision mode, we first select a random radius $r_p$ in the interval $[r_{t}-\delta, r_{t})$ and $(r_{t}, r_{t}+\delta]$. $(r_t-\delta \geq 0\ \&\&\ r_t+\delta \le r_{max})$. We can generate an image $\mathcal{A}(x_{i}, r_p)$. Given the image $\mathcal{A}(x_{i}, r_p)$, our backdoored model should not trigger the backdoor but return the correct class prediction. We can use the training set with precision mode to refit $\Theta^{adv}$.\par

\begin{equation}
(x_{i},y_{i})\mapsto
\begin{cases}
(x_{i}, y_{i}) & \text { \textbf{Clean mode}}  \\
(\mathcal{A}(x_{i}, r_{t}), t) & \text { \textbf{Attack mode}} \\ 
(\mathcal{A}(x_{i}, r_p), y_{i}) & \text {\textbf{Precision mode}} 
\end{cases}
\label{Equation:precision_loss}
\end{equation}

\subsection*{Backdoor injection}
After generating poisoned dataset $D_{poisoned}$, clean dataset $D_{clean}$ and precision dataset $D_{precision}$ according to three modes, we can retrain the model parameters $\Theta^{adv}:=\Theta^{'}$. We conclude our low-pass attack in algorithm~\ref{algorithm:Low-pass_attack_precision}.
In the inference phase, we apply the same trigger radius: $r_{t}$ used in the training phase to generate poisoned validation samples. We will show our experiment results in the next section. 
\begin{table}[h!]
\caption{Attack performance on different datasets. $r_{max}$ denotes the max low-pass radius of the dataset $(0<r_t<r_{max})$. "Best $r_t$" denotes the max $r_t$ without compromising CSA and ASR.}
  \begin{tabular}{cccccc}
    \hline
    Dataset & Input Size & $r_{max}$&CSA & ASR & Best $r_{t}$ \\
    \hline
    MNIST & $28\times28$ & 14 & 99.37 & 99.78 &12\\
    CIFAR10 &$3\times32\times32$ & 16 &  87.74& 97.10 &12\\
    GTSRB & $3\times32\times32$ & 16 &98.47& 98.27 &13\\
    CelebA & $3\times64\times64$ & 32 &77.92& 99.62 &25\\
    \hline
  \end{tabular}
  \label{tab:dataset_asr}
\end{table}

\section*{Experiments and Analysis}

In this section, we implement our backdoor attack introduced in Section 3. \par
For our low-pass based backdoor attack, we mount our attack on MNIST~\cite{lecun1998mnist}, CIFAR-10~\cite{krizhevsky2009learning}, GTSRB~\cite{stallkamp2012man} and CelebA~\cite{liu2015deep}. Note that in CelebA, we select the top three most balanced attributes, including Heavy Makeup, Mouth Slightly Open, and Smiling, then concatenate them to create eight classification classes. All datasets are widely used in deep learning. Our experiments were run on a machine with two 2080Ti, and our networks were implemented by Pytorch 1.5~\cite{paszke2019pytorch}.
For MNIST, we use AlexNet~\cite{krizhevsky2017imagenet} as our baseline model. For CIFAR10, GTSRB, and CelebA, we use pre-trained ResNet-18~\cite{he2016deep} as the original model. We trained the networks using SGD~\cite{ruder2016overview} optimizer until convergence. The initial learning rate was 0.01 and was reduced by a factor of 10 after 100 epochs.\par
The success of a backdoor attack can be generally evaluated by Clean Sample Accuracy(CSA) and Attack Success Rate(ASR), which can be defined as follows:\\ 
\textbf{Clean Sample Accuracy (CSA):} For normal users, the CSA measures the proportion of clean test samples containing no trigger that is correctly predicted to their ground-truth classes. \\
\textbf{Attack Success Rate (ASR):} For an attacker, we represent the output of the backdoored model $M_{{\Theta^{adv}}}$ on poisoned input data $x^{poisoned}$ as $y^{'}= M_{{\Theta^{adv}}}(x^{poisoned})$ and the attacker's expected target as $t$. This index measures the ratio of $y^{'}$ which equals the attacker target $t$. This measurement also shows whether the neural network can identify the trigger pattern added to the input images.\\
\textbf{Pollution Rate (PR):} It is defined as the fraction of the poisoned training dataset.

\begin{figure}
  \centering
  \includegraphics[width=0.96\linewidth]{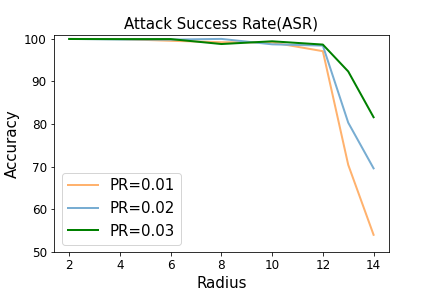}
  \caption{The relationship of the attack success rate(ASR) and attacker chosen trigger radius: $r_t$ under different pollution rate (PR) on CIFAR10. Note that CSA remains around 87\% for various $r_{t}$ and PR.}%
  \label{fig:pr_csa_asr}
\end{figure}

\subsection*{Low-pass based backdoor attack}

To demonstrate the feasibility of our low-pass attack, we implement algorithm~\ref{algorithm:Low-pass_attack_precision}. We apply different low-pass trigger radiuses: $r_t$ and different pollution rates (PR) to train multiple backdoored models. When validating the backdoored model, we apply the same low-pass trigger radius: $r_t$ set in the training phase on original validation dataset, and then we record the attack performance. \par
First, we demonstrate the feasibility of our attack on CIFAR10. Note that $r_{max}$ is defined as half of the image height(or width). For example, the shape of each image in CIFAR10 is $3\times32\times32$, the max low-pass radius is $r_{max}=16$. We valid the attack performance when changing trigger radius: $r_t$ and pollution rate (PR=0.01, 0.02, 0.03). As Figure~\ref{fig:pr_csa_asr} shows, the attack success rate is close to $100\%$ for any low-pass radius: $r_t$ less than or equal to 12, even if the pollution rate (PR) is 0.01. Our clean sample accuracy (CSA) remains almost constant (around $87\%$) for different $r_t$ and different PR.\par
\begin{figure*}[tbp]
    \centering
    \includegraphics[width=2\linewidth]{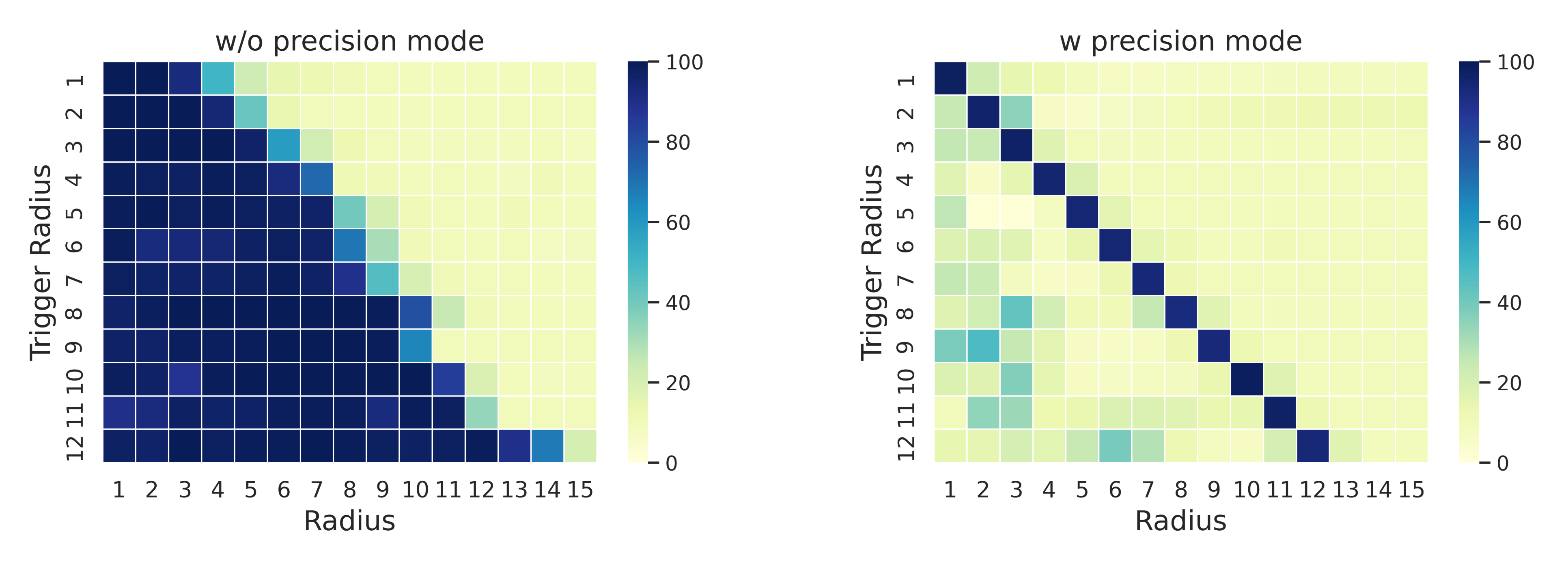}
    \caption{Attack Success Rate(ASR) of images radiuses with and without precision mode.
    Note that we train 12 backdoored models according to 12 trigger radiuses: $r_{t}$ and we valid the ASR of images with all the radiuses under the 12 backdoored models. Y axis indicates various trigger radiuses: $r_{t}$ selected by the attacker. X axis represents all the radiuses. Color depth in the heat map represents the ASR. Precision mode serves to constrain the ASR of other radiuses: $r^{'}$($r^{'}\neq r_{t}$) while ensuring the ASR of the trigger radius: $r_{t}$.}%
    \label{fig:precision_mode}
\end{figure*}
We also perform extensive experiments on MNIST, GTSRB, and CelebA. Table~\ref{tab:dataset_asr} shows our attack performance on different datasets. Our pollution rates (PR) are all set to 0.01. In the table, the "Best $r_t$" index denotes the max $r_t$ without compromising ASR and CSA (a larger $r_t$ preserves more semantic information of the original image.). We aim to preserve more information of original images while ensuring CSA and ASR.

\subsection*{Ablation study}
In order to show the role of our precision mode, we trained various backdoored models according to different trigger radiuses: $r_{t}$ with and without precision mode on CIFAR10. Note that we all chose $\beta=0.01$ as pollution rate. In the inference phase, we record the attack success rate (ASR) of images with all the radiuses. Trigger radius neighborhood: $\delta$ is a hyper-parameter that can be adjusted. For different trigger radiuses: $r_{t}$, we apply different $\delta$ to achieve better results. From figure~\ref{fig:precision_mode}, we can observe that though we can achieve a perfect attack success rate without precision mode, our attack is not that precise as the backdoor can be triggered as long as the low-pass radius: $r^{\prime}$ is within a certain range. When we apply precision mode as a penalty term, the images with other low-pass radius: $r^{\prime}$ will lose their ability to attack the model. The backdoor can only be precisely triggered by our specified low-pass radius: $r_t$. Note that the use of precision mode can decrease the attack success rate of $r_t$ slightly.

\subsection*{Discussion on defense methods}
\begin{figure*}[tbp]
    \centering
    \includegraphics[width=2\linewidth]{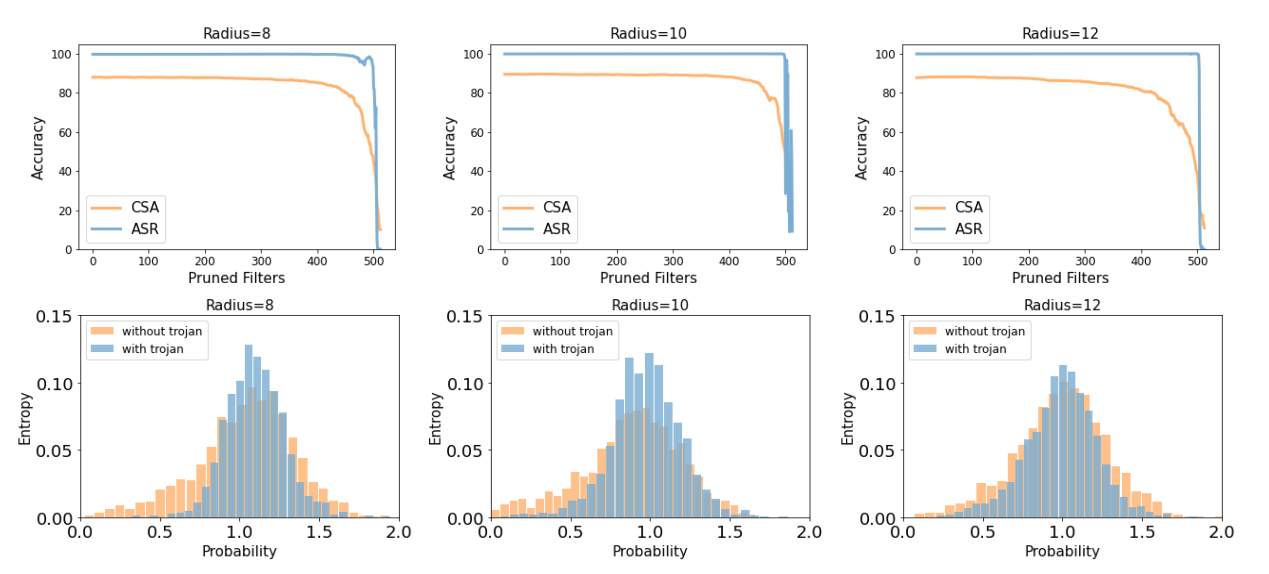}
    \caption{Experiments on verifying low-pass attack by Fine-pruning and STRIP. Note that on CIFAR10, we train three backdoored models according to three different trigger radiuses: $r_t=8,10,12$. Row 1, row 2 represent the results of our attack against Fine-pruning and STRIP respectively.}%
    \label{fig:strip_fine_pruning}
\end{figure*}

In this section, we conduct several experiments to prove the effectiveness of our attack against three popular defense methods, including STRIP, Fine-pruning, and Neural Cleanse. Note that the larger the trigger radius, the better the invisibility of the poisoned images. Therefore, to ensure both the invisibility and ASR, we apply three larger trigger radiuses: $r_{t}=8, 10, 12$ to train three backdoored models as examples.\\
\textbf{STRIP} is a test-time defense method. It takes advantage of the feature that any input containing a backdoor will be classified as the target (if the model contains a backdoor, the input data containing the backdoor will be classified as the specified target after superimposing, while the classification result of superimposed clean input is relatively random). It distinguishes clean inputs from poisoned inputs by determining the information entropy of the model output classification results. With our low-pass based backdoor attack, the perturbation will modify our poisoned image content. As shown in the second row of Figure~\ref{fig:strip_fine_pruning}, STRIP cannot distinguish poisoned inputs from clean inputs. \\
\textbf{Fine-pruning} removes backdoor by performing pruning operations on the backdoored model. Given a clean dataset, defenders assume that the dormant neurons can only be activated by poisoned inputs. Therefore, they detect dormant neurons from a specified layer and then remove the backdoor by pruning these dormant neurons. We apply the fine-pruning technique to prune our trained low-pass based model. As shown in the figure, ASR only starts to drop when CSA drops to about 40\%, indicating that this defense method cannot effectively eliminate our backdoor attack.\\
\textbf{Neural cleanse} uses an optimization method to perform model defense. Assuming the backdoor trigger is patch-based, for each class label, this method computes the most stealthy patch pattern in the premise of converting any clean inputs to the target label. After that, it checks whether any label has a significantly smaller pattern as a sign of a backdoor. Neural Cleanse determines whether the model has a backdoor by an Anomaly Index with threshold $\tau=2$. We ran the defense on our three backdoored models and recorded the results in Table \ref{tab:neaural_cleanse}. From the results, we can see that the anomaly indices are all less than the set threshold value.
\begin{table}[h!]
\caption{Experiments on verifying low-pass attack by Neural cleanse. On CIFAR10, we train three backdoored models according to three different low-pass radiuses: $r_{t}=8,10,12$.}
  \begin{tabular}{cccc}
    \hline
    Radius: $r_t$ & Median & MAD & Anomaly index \\
    \hline
    8 &  40.30 & 3.16 & \textbf{1.51}\\
    10 &  35.30& 6.63 & \textbf{0.67}\\
    12 & 40.73& 7.68 & \textbf{1.22}\\
    \hline
  \end{tabular}
    \label{tab:neaural_cleanse}
\end{table}

\subsection*{Comparison with other attacks}
Here, we compare our low-pass attack with existing backdoor attacks including BadNets~\cite{gu2017badnets}, SIG~\cite{barni2019new} and Refool~\cite{liu2020reflection}. For BadNets~\cite{gu2017badnets}, we add a $4\times4$ white block in the bottom right corner of original images as a trigger. We apply two metrics to measure the invisibility of different methods. PSNR~\cite{huynh2008scope} measures the ratio of the maximum pixel value of an image to the mean squared error (between clean and poisoned images.) A larger PSNR means that the perceived difference between the two images is smaller. SSIM~\cite{wang2004image} is an index that measures the similarity of two images. It is calculated based on the brightness and contrast of the local patterns. the larger the SSIM, the better the quality of the poisoned image. From table~\ref{tab:compare_cifar10}, we can observe that our attack outperforms the competitors for the two invisibility metrics without compromising attack performance.

\section*{Conclusion}
In this paper, we propose a novel and effective backdoor attack based on image low-pass filter. Unlike existing poisoned image generation approaches, our method reduces high-frequency components and preserves the semantic integrity of original images rather than adding extra perturbations, enhancing the capacity to evade current defenses. In addition, we introduce "precision mode" into our training procedure, making the backdoor triggered more precisely and stealthily. Experimental results prove that the proposed method generates effective adversarial backdoor images and can evade various state-of-the-art defences. The poisoned images of our low-pass attack also retain higher image quality compared with existing attacks. Our low-pass based method opens a new area of backdoor attack and encourages future defense research.

\begin{table}[h!]
\caption{Comparison with other attacks on CIFAR10. Note that we select attack radius: $r_t=12$ to generate poisoned samples.}
  \begin{tabular}{ccccc}
    \hline
    Methods & CSA & ASR & PSNR & SSIM \\
    \hline
    BadNets & 87.01 & 96.36 & 23.4 & 0.937 \\
    SIG &  86.47 & 96.42 & 26.2 & 0.862 \\
    Refool & 86.91 & 75.20 & 17.1 & 0.778 \\
    Ours & \textbf{87.74} & \textbf{97.10} & \textbf{32.9}&\textbf{0.983} \\
    \hline
  \end{tabular}
\label{tab:compare_cifar10}
\end{table}

\bibliographystyle{bmc-mathphys} 
\bibliography{bmc_article} 
\end{document}